\documentclass[11pt,twoside]{article}

\usepackage{asp2006}
\usepackage{epsf}
\usepackage{epsfig}
\usepackage{lscape}

\newcommand{\zphot}{\ensuremath{z_{\rm phot}} }
\newcommand{\zspec}{\ensuremath{z_{\rm spec}} }

\newcommand{\cl}{Cl~0332-2742}

\begin{document}
\title{A spectroscopic study of a $z=1.6$ galaxy overdensity in the GMASS field}   
\author{Jaron Kurk\altaffilmark{1,2}, 
Andrea Cimatti\altaffilmark{3,2},
Gianni Zamorani\altaffilmark{3},
Claire Halliday\altaffilmark{2},
Marco Mignoli\altaffilmark{3},
Lucia Pozzetti\altaffilmark{3},
Emanuele Daddi\altaffilmark{3},
Piero Rosati\altaffilmark{3},
Marc Dickinson\altaffilmark{3},
Micol Bolzonella\altaffilmark{3},
Paolo Cassata\altaffilmark{3},
Alvio Renzini\altaffilmark{3},
Alberto Franceschini\altaffilmark{3},
Giulia Rodighiero\altaffilmark{3},
Stefano Berta\altaffilmark{3}}   
\altaffiltext{1}{Max-Planck-Institut f\"ur Astronomie, K\"onigstuhl 17, 
                 D-69117, Heidelberg}
\altaffiltext{2}{INAF-Osservatorio Astrofisico di Arcetri, 
                 Largo E. Fermi 5, I-50125, Firenze}
\altaffiltext{4}{Other institute: Universit\'a di Bologna, INAF-Bologna,
                 INAF-Bologna, INAF-Bologna, CEA-Saclay, ESO-Garching
                 bei M\"unchen, NOAO-Tucson, INAF-Bologna,
                 OAMP-Marseille, Universit\'a di Padova, INAF-Padova,
                 INAF-Padova, MPE-Garching bei M\"unchen,
                 respectively.}


\begin{abstract} 
  The Galaxy Mass Assembly ultra-deep Spectroscopic Survey samples a
  part of the CDFS to unprecedented depth. The resulting distribution
  of 150 $z>1.4$ redshifts reveals a significant peak at $z = 1.6$,
  part of a larger overdensity found at this redshift. The 42
  spectroscopic members of this structure, called Cl~0332-2742, form
  an overdensity in redshift of a factor 11$\pm$3 and have a velocity
  dispersion of 450 km\,s$^{-1}$.  We derive a total mass for
  Cl~0332-2742 of $\sim 7 \times 10^{14}$ M$_\odot$.  The colours of
  its early-type galaxies are consistent with a theoretical red
  sequence of galaxies with stars formed at $z=3.0$.  In addition,
  there are more massive, passive and older, but less star forming
  galaxies in CL~0332-2742 than in the field. We conclude that this
  structure is a cluster under assembly at $z=1.6$.
\end{abstract}



\section{Introduction}   

It has been known for a considerable time that the environment plays
an important role in galaxy evolution.  Observations of low and high
redshift clusters have provided evidence that the fraction of blue and
red galaxies changes considerably between $z = 0$ and $z \sim 0.5$
\citep[e.g., ][]{but84}. In addition, the galaxy population inside
clusters evolves at a rate different from that in the field \citep[e.g.,
][]{and06, tra07, fas08}.  
Distant galaxy clusters, therefore, provide important environments for
the study of galaxy evolution, especially if both passive and active
galaxy populations can be studied in detail.  For the known high
redshift overdensities this is often problematic, because their
cluster members are selected either on their star formation activity
\citep[blue members, e.g.,][]{ven02, ste05}, or lack of
it \citep[red members, e.g.,][]{kod07, mcc07}.  The best
studied, established, high redshift clusters are found between $z =
1.0$ and $z = 1.4$ 
\citep{sta05,sta06}.  We present
a galaxy overdensity at $z = 1.61$, which appears to be a cluster
under assembly, containing both red and blue galaxy populations.

\section{Data}   

GMASS ({\it Galaxy Mass Assembly ultra-deep Spectroscopic Survey}) is
a project based on an ESO VLT Large Program.  To obtain a mostly mass
selected sample, we extracted all sources present in the $4.5\,\mu$m
Spitzer/IRAC image to a limiting magnitude of $m_{4.5}<23.0$
(2.3\,$\mu$Jy), in a region of $6.8\times6.8$ arcmin$^2$ located
within GOODS-South. The GMASS sample includes 1277 unblended objects
to $m_{4.5}<23.0$, with photometry from the NUV to MIR and SED fits
with \citet[][ M05]{mar05} templates providing stellar masses, star
formation rates and other galaxy properties.  The VLT/FORS2 optical
spectroscopy was focused on galaxies pre--selected using a cut in
photometric redshift of $z_{\rm phot}>1.4$ and two cuts in the optical
magnitudes ($B,I<26.5$).  This selection resulted in 221 spectroscopic
targets, 170 of which were actually observed.  The integration times
were very long (up to 32 hours per mask and some targets were included
in multiple masks), and the spectroscopy was optimized by obtaining
spectra in the blue (4000--6000\,\AA) or in the red (6000--10000\,\AA)
depending on the colours and photometric redshifts of the targets.
Despite the faintness of the targets, GMASS spectroscopy was very
successful and provided an overall spectroscopic redshift success rate
of about 85\%.

\begin{figure}
  \label{fig:fig1}
  \epsfig{file=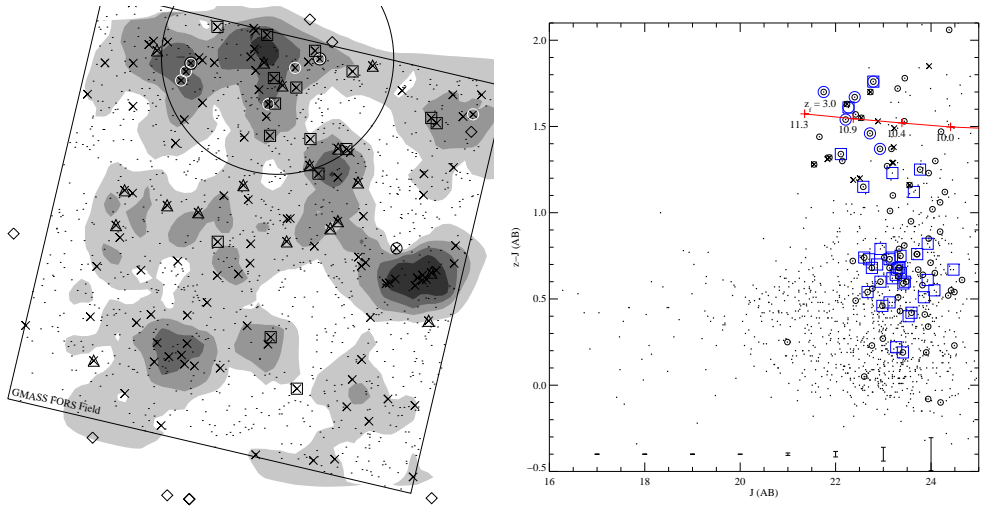,width=\linewidth,clip=}
  \caption{{\bf Left} Contours of $z = 1.6$ galaxy density as
    described in the text, linearly increasing up to 3.4 times the
    median density.  Candidate members are indicated by crosses,
    spectroscopic members by circles (for elliptical), squares (for
    spiral) and triangles (for irregular galaxies).  Early, late and
    intermediate types are indicated by white, black and white-black
    symbols.  Diamonds indicate members not in the GMASS catalog. {\bf
      Right} Colour-magnitude ($z-J$ vs $J$) diagram for the entire
    GMASS catalog.  Galaxies with $1.600 < \zspec < 1.622$ are
    indicated by large symbols, and with $1.50 < \zphot < 1.70$ by
    small circles, except if outside $1.600 < z_{\rm spec} < 1.622$
    (crosses).  The large symbols are circular for early types, and
    square for late types.  The (red) line indicates a theoretical red
    sequence of elliptical galaxies, observed at $z=1.60$, with masses
    indicated in logarithmic solar masses, formed at $z_{\rm f} =
    3.0$.  Typical errors in colour are also indicated at the 
    bottom.\vspace{-1.2cm}}
\end{figure}

\section{Results}   

Several peaks in the distribution of GMASS redshifts (at $z>1.4$)
stand out, the peak at $z = 1.6$ being the most significant at $z >
1.4$.  Although this peak was known to exist
\citep{gil03,van05,van06,cim02,cim04,cim08} and described in detail by
\citet{cas07}, the availability of 136 spectroscopic redshifts at $z >
1.4$ from GMASS allows a far better study than before.  We call this
redshift peak \cl.


The angular distribution of the $z = 1.6$ galaxies in the GMASS field
is shown in Fig.~1.  Here we have used both spectroscopic information,
showing members with $1.600 < \zspec < 1.622$ and the photometric
redshifts, showing candidate members with $1.43 < \zphot < 1.77$,
except for galaxies with $\zspec < 1.600$ or $\zspec > 1.622$.  The
associated surface density contours show that there are two localized
overdensities: one towards the northern edge and one towards the
western edge of the GMASS field.  This is a striking coincidence of
high density and galaxy type, consistent with density relations found
at lower redshift.  To make these statements more quantitative, we
have defined a circular \emph{high density region} (see Fig.~1),
following \citet{cas07}, centered on a remarkably large galaxy in the
centre of the northern overdensity, and with a radius of one
(physical) Mpc.  The \emph{high--density region} contains 21 of the
galaxies with $\zspec = 1.6$; a similar number is present in the area
outside this region, which is, however, five times larger.  From the
six early and intermediate type-galaxies among the members, five are
within the high--density region, while three more, west of the GMASS
field, are known from the K20 survey \citep{cim04}.

To compare the properties of the member galaxies with field galaxies,
we select a sample of 43 galaxies outside the peak, i.e., 24 in the
redshift interval $1.416 < z < 1.598$ and 19 in the interval $1.624 <
z < 1.840$.  The field galaxies were selected within the GMASS field
and have therefore similar, identically derived, information available
to the galaxies in \cl.  There are significant differences (KS test
probabilities $<$1\%) for many physical properties between the
galaxies in the high--density region and the field.
The brightest confirmed galaxy member is located, together with two
other bright galaxies in the north-east of the GMASS field. It is
remarkable that this triplet of massive, passive and red galaxies lies
about 1.5\arcmin\ (760\,kpc) eastward of the centre of the high
density region.  This may be an indication that the cluster is not 
relaxed.

Employing the biweight statistic, we estimate a velocity dispersion
for the 42 galaxies of $\sigma = 440^{+95}_{-60}$ km\,s$^{-1}$ and
median redshift $z_{\rm med} = 1.610$.  Considering only the 21
galaxies within the high density region, we obtain a velocity
dispersion of $\sigma_{\rm high} = 500^{+100}_{-100}$ km\,s$^{-1}$.
The distribution of the 42 redshifts is not consistent with a Gaussian
distribution.  Its asymmetry may be an indication that \cl\ is not
relaxed.  Although we do not know whether the structure under study
here is virialized, we apply the usual relations for virial radius and
mass, obtaining estimates of $R_{200}$ = 0.5 Mpc and $M_{\rm vir}$ = 9
$\times 10^{13}$ M$_\odot$.


To estimate the galaxy overdensity in redshift, we assume a flat
n$(z)$ distribution and employ the number of 108 galaxies with
spectroscopically--confirmed redshifts within the GMASS sample in the
range $1.400 < z < 1.900$.  We find an overdensity of 11$\pm$3, where
the uncertainty is determined by excluding the $z=1.6$ galaxies in the
flat n$(z)$ distribution or not.  Using this overdensity and the
volume defined by the GMASS field size and the redshift interval
$1.600 < z < 1.616$, we obtain a total mass within \cl\ of $6.7\pm1.6
\times 10^{14} $M$_\odot$.  Considering only the volume occupied by
the high density region, and its overdensity, we obtain a total mass
of $7 \times 10^{13} $M$_\odot$.

The field population of galaxies up to at least $z \sim 1$ shows a
bimodal distribution in colour \citep[e.g.,][]{bel04b}.  Within
cluster populations, the bimodality is detected as a concentration of
red (early--type) galaxies on a line in a colour-magnitude diagram
(CMD) using colours which straddle the 4000\,\AA\ break.  For $z =
1.6$, the break is shifted to 1.04\,$\mu$m and a suitable CMD is
therefore a plot of $z-J$ vs $J$ (Fig.~1).  There appear to be two
populations of spectroscopic members: a blue population around $z-J =
0.6$, and a red population with $z-J \sim 1.5$.  Clearly the early and
intermediate types (derived from the spectra) have the reddest $z-J$
colours: all are at $z-J > 1.3$.  We overplot a line that represents a
theoretical red sequence at $z = 1.60$, corresponding to the predicted
$z$ and $J$ magnitudes of the stellar populations of elliptical
galaxies that formed in a burst of 0.5 Gyr at $z_{\rm f} = 3.0$
\citep{kod98}.  This line appears to represent very well the sequence
formed by the red galaxies in \cl.

The structure at $z=1.6$, described here, is most probably a galaxy
cluster progenitor: there is a galaxy overdensity of at least a factor
eight, a red sequence of passive galaxies, and evidence that the
galaxies in the structure are significantly more evolved and more
massive than field galaxies.  However, the irregularity of the
structure and the weak X-ray emission, suggest that the cluster is not
relaxed yet.  These results appear to be consistent with predictions
of models of hierarchical galaxy evolution.  We conclude that \cl\ is
a galaxy cluster under assembly (see Kurk et al., submitted, for more
details).

\acknowledgements 

JDK is supported by DFG/SFB-439.  This work is based on observations
made with ESO's VLT (LP 173.A-0687) and with the Spitzer Space
Telescope, operated by JPL, CIoT under a contract with NASA.





\end{document}